\documentclass[%
 reprint,
nofootinbib,
 amsmath,amssymb,
 aps,
]{revtex4-2}

\usepackage{graphicx}% Include figure files
\usepackage{dcolumn}% Align table columns on decimal point
\usepackage{bm}% bold math
\usepackage{multirow}
\usepackage{amssymb,mathrsfs,amsmath}
\usepackage{amsfonts}
\usepackage{color,xcolor}
\usepackage{subfigure}
\def\linkcolor{cyan!70!black}

\usepackage[
colorlinks=true
,urlcolor=\linkcolor
,anchorcolor=\linkcolor
,citecolor=\linkcolor
,filecolor=\linkcolor
,linkcolor=\linkcolor
,menucolor=\linkcolor
,linktocpage=true
,pdfproducer=medialab
,pdfa=true
]{hyperref}

\newcommand{\bc}{\begin{center}}
\newcommand{\ec}{\end{center}}
\newcommand{\be}{\begin{equation}}
\newcommand{\ee}{\end{equation}}
\newcommand{\ba}{\begin{eqnarray}}
\newcommand{\ea}{\end{eqnarray}}
\newcommand{\bea}{\begin{eqnarray*}}
\newcommand{\eea}{\end{eqnarray*}}

\begin{document}

\preprint{FT-HEP24-2}
\title{ The $\rho$ meson magnetic dipole moment from BaBar $e^+e^- \to \pi^+ \pi^-\pi^0 \pi^0 $ cross section measurement}% Force line breaks with \\

\author{Antonio Rojas}
\affiliation{Instituto de F\'{\i}sica,  Universidad Nacional Aut\'onoma de M\'exico, AP 20-364,  M\'exico D.F. 01000, M\'exico}%
\author{Genaro Toledo}
\affiliation{Instituto de F\'{\i}sica,  Universidad Nacional Aut\'onoma de M\'exico, AP 20-364,  M\'exico D.F. 01000, M\'exico}%

\date{\today}% It is always \today, today,
             %  but any date may be explicitly specified

\begin{abstract}
We obtain the value of the magnetic dipole moment of the $\rho$ meson, using data from the BaBar Collaboration for the $e^{+} e^{-} \to \pi^+ \pi^- 2 \pi^0$ process. The considered center of mass energy range is from 0.9 to 1.8 GeV. We describe the $\gamma^* \to 4\pi$ vertex using a vector meson dominance model, including the intermediate resonant contributions relevant at these energies. We find $\mu_\rho =  2.7 \pm 0.3 $ in $e/2 m_\rho$ units. We improve on the previous extracted value, where preliminary data from the same collaboration was used, by considering definite data, better grounded values of the parameters involved and explicit gauge invariant description of the process.
\end{abstract}

%\keywords{rho meson; magnetic dipole moment}
%Use showkeys class option if keyword
                              %display desired
\maketitle
\section{Introduction}
The magnetic dipole moment (MDM) has played an important role in the understanding of the dynamics of elementary particles. The gyromagnetic ratio, $g$, defines the value of the MDM in units of the magneton of the particle ($e/2M$). Since the early work of Dirac, it has been  observed that g=2 is valid not only for spin one-half but also for higher spin particles \cite{Deser:1970spa,Brodsky:1992px,Ferrara:1992yc, Jackiw:1997vw,ToledoSanchez:2002rn, Holstein:2006wi,Dbeyssi:2011ep,Lorce:2009bs,Delgado-Acosta:2012dxv,Haberzettl:2019qpa}. This value is commonly used as a reference for structured states, where corrections from the internal dynamics are expected.  

The $\rho$ vector meson is one of such states, whose MDM has been predicted in different approximations to QCD. Relativistic quark models (RQM) including light-front dynamics and sum rules
 \cite{Bagdasaryan:1984kz,Cardarelli:1994yq, deMelo:1997hh, Melikhov:2001pm,Alexander_2003,Aliev:2003ba,Jaus:2002sv,Choi:2004ww,He:2004ba,ALIEV2009470,Biernat:2014dea,Simonis2016,deMelo:2016ynt,Krutov:2018mbu,Simonis:2018rld,DeMelo:2018bim}, 
Dyson-Schwinger equations
(DSE) \cite{Hawes:1998bz, Bhagwat:2006pu,Roberts:2011wy, Pitschmann:2012by, Xu:2019ilh, Xing:2021dwe,Shi:2023oll,Xu:2023vlt}, 
Poincar\'e-Invariant Quantum mechanics (PIQM), \cite{Haurysh2021}, 
Nambu-Jona-Lasinio model (NJL) \cite{Zhang:2022zim, Luan_2015},
Chiral effective field theory for vector mesons \cite{Djukanovic:2013mka},
Lattice QCD \cite{Andersen:1996qb, Dudek:2006ej, Hedditch:2007ex,Lee:2008qf, Shultz:2015pfa,Owen:2015gva, Luschevskaya:2016epp}.
According to such predictions, the MDM value lies in the range $\mu_\rho =[1.5 \, , \, 2.7] $ in $e/2 m_\rho$ units. 

An experimental counter part to confront these predictions has proven to be elusive, given the instability feature of the $\rho$, which prevents experimentalists from applying standard techniques to measure the MDM. For example, the one used for the muon MDM, which now reaches unprecedented precision \cite{Muong-2:2023cdq}, requires to store the particle for a relatively long time in presence of a magnetic field \cite{Bargmann:1959gz}. Alternatives to surpass this difficulty have been proposed. Radiative decays, where the photon emission can be split into the different multipoles offers another possibility \cite{Zakharov:1968fb}. Processes such as $\tau\to \pi\pi\nu_\tau\gamma$ \cite{LopezCastro:1999dp,LopezCastro:1999xg} and $\rho\to \pi\pi\gamma $ \cite{LopezCastro:1997dg}, have shown to be sensitive to the $\rho$ MDM. The finite decay width has been also incorporated \cite{LopezCastro:1999xg} and the impact on the multipoles shown to be very mild \cite{GarciaGudino:2010sd}. However, there is no experimental data available to extract a value for the MDM. This approach has been successful in the case of the $\Delta^{++}$, where the MDM has been determined \cite{LOPEZCASTRO2001339}. 

Another approach is the one inspired in the procedure followed to extract the $W$ gauge boson MDM \cite{DELPHICollab,MELE2004255}. There, the $e^+ e^- \to jjl\nu$ process is considered (where $j$, $l$ and $\nu$ denotes a jet, a lepton and a neutrino, respectively), which involves the $WW\gamma$ vertex at the intermediate stage. The key idea is that such vertex carries information on the multipole structure of the radiating particle. This suggests that, for the $\rho$ meson, the $e^{+} e^{-} \to \pi^+ \pi^- 2 \pi^0$ process, which involves the $\rho\rho\gamma $ vertex, would be an ideal scenario to determine the MDM. The $e^{+} e^{-} \to \pi^+ \pi^- 2 \pi^0$ process is within current experimental capabilities. It has been measured by several experiments in the low energy regime, using a direct production mechanism \cite{Cosme:1978qe,Bacci:1980zs, Kurdadze:1986tc,DM2:1990npw, Dolinsky:1991vq,CMD-2:1998gab,Achasov:2003bv, Achasov:2009zz}. The BaBar collaboration measured the process in a wider energy range, using the initial state radiation technique. A preliminary data from BaBar \cite{druzhinin2007, Druzhinin:2011qd} was used in
a previous work \cite{GarciaGudino:2013alv,David2015}, to explore this idea. A value of $\mu_\rho =  2.1 \pm 0.5 $ in $e/2 m_\rho$ units was obtained, based on a vector meson dominance (VMD) description of  the total cross section \cite{Kroll:1967it,Bando:1984ej, Fujiwara:1984mp, Meissner:1987ge}. The analysis took into account the preliminary status of the data, the uncertainties coming from the model dependence and the parameters involved. A VMD-like ansatz for the $\rho^\prime$ parameters was invoked. The result provided a first value driven by experimental data, although with large error bar. This result has been considered in new theoretical and experimental studies, which welcomed the possibility of comparing with a data driven value \cite{Gegelia:2014tma, Biernat:2014dea,Owen:2015gva,Carrillo-Serrano:2015uca,Simonis2016,Luschevskaya:2016epp,Solovjeva:2016fxy,Sun:2017gtz,deMelo:2016ynt, DeMelo:2018bim,Simonis:2018rld,Andreichikov:2019xmm, Sun:2018ldr,Aliev:2019lsd,Xu:2019ilh,Meng:2022ozq,Hernandez-Tome:2023ujh}.\\
Definite experimental information has been released by the BaBar Collaboration on the same process \cite{BaBar:2017zmc}. It also includes exclusive data on the $\omega$ channel, a subprocess relevant in the low energy region. This can be used to constrain the model parameters in a more robust way. Thus, a reanalysis of the process is in order, to obtain a better grounded value of the $\rho$ MDM consistent with the BaBar data.\\
In this work, we obtain the $\rho$ MDM following the same approach as in the previous analysis \cite{GarciaGudino:2013alv,David2015}, while improving at several stages, as we mention below. In Section II, we present the general features of the electromagnetic vertex to set the context of the analysis. Then, we describe the $e^{+} e^{-} \to \pi^+ \pi^- 2 \pi^0$ process, by modeling the $\gamma^* \to 4\pi$ vertex in the VMD approach \cite{Kroll:1967it,Bando:1984ej, Fujiwara:1984mp, Meissner:1987ge}. This description allows to include the relevant hadronic degrees of freedom in the energy range of our interest, provided the effective coupling constants are known. 
There, we obtained the explicit gauge invariant amplitude of the process, paying particular attention to the channels  linked by gauge invariance, and involving the $\rho$ MDM.
Bose-Einstein symmetry and charge conjugation (C) invariance are enforced by properly including the exchange of neutral and charged pions, respectively. We profit from a previous analysis of the description of low energy observables to fix the parameters of the model \cite{Avalos:2022txl}. We constrain them further, by incorporating the data of the so-called $\omega$ channel from BaBar \cite{BaBar:2017zmc}. As a complementary check of the parameters involved, we show that, in the low energy regime, the BaBar data is well described. In that regime, the $\rho$ MDM plays no role.
In Section III, we exhibit the total cross section dependence on the $\rho$ MDM. We determine its value by fitting the experimental data, while using other observables to fix all the remaining parameters. 
In Section IV, we discuss the result and draw our conclusions.

\section{Description of the $e^{+} e^{-} \to \pi^+ \pi^- 2 \pi^0$  process}

For a vector particle $V (q_i,\eta_i)$, the photon radiation process $V(q_1, \eta_1) \to V(q_2, \eta_2) \gamma (q,\epsilon)$, where in parenthesis are the corresponding momentum and polarization vector, is defined from the matrix element
\begin{equation}
    \langle V(q_2, \eta_2)|J^\mu_{EM}(0)|V(q_1, \eta_1) \rangle \equiv e\Gamma^{\mu \nu \lambda} \eta_{1\nu} \eta^*_{2\lambda},
\end{equation}
The C, P (parity) and CP conserving electromagnetic vertex $\Gamma^{\mu \nu \lambda}$ can be decomposed into the following Lorentz structures 
\begin{eqnarray}
\Gamma^{\mu\nu \lambda} &=&
 \alpha(q^{2}) g^{\nu \lambda}(q_1 + q_2)^{\mu} + \beta(q^{2}) ( g^{\mu \nu} q^{\lambda} -  g^{\mu \lambda} q^{\nu})\nonumber\\
  &-& \frac{\gamma(q^{2})}{M_V^2} (q_1 + q_2)^{\mu} q^{\nu} q^{\lambda} 
  - q_1^\lambda g^{\mu \nu} -q_2^\nu g^{\mu \lambda},
\label{vertex}
\end{eqnarray}
where $\alpha(q^{2})$, $\beta(q^{2})$  and $\gamma(q^{2})$ are the electromagnetic form factors \cite{Kim:1973ee,Hagiwara:1986vm,Nieves:1996ff,Gounaris:1996rz}. In the static limit, they are related to the electromagnetic multipoles as follows: $\mathcal{Q}_V = \alpha(0)$ is the electric charge (in $e$ units), $\mu_V=\beta(0)$ is the magnetic dipole moment (in $e/2 M_V$ units)  and $X_{E_V} = 1-\beta(0)+2\gamma(0) $ (in $e/M_V^{2}$ units) is the electric quadrupole moment. Other parameterization is the one used for the $W$ gauge boson, given in terms of the parameters $\kappa_{\gamma}$ and $\lambda_{\gamma}$, which are related to the multipoles by:  $\mu_W =\frac{e}{2\,m_W}\,(1+\kappa_{\gamma}+\lambda_{\gamma})$ and $Q_W =-\frac{e}{m^{2}_W}\,(\kappa_{\gamma}+\lambda_{\gamma})$.
 The values $\alpha(0) = 1$, $\beta(0) = 2$ and  $\gamma(0) =0 $ ($\kappa_{\gamma}=1$ and $\lambda_{\gamma}=0$ for the $W$) are usually taken as reference for vector mesons. For the $\rho$ meson, the physical value is expected to be affected by the strong interaction dynamics among quarks. Notice that the MDM grows as ${\cal O}(q)$ while the quadrupole does it as  ${\cal O}(q^2)$. Thus, the MDM is expected to be dominant for a relatively small photon energy, compared to the quadrupole moment. This behavior was verified in the previous analysis \cite{GarciaGudino:2013alv,David2015}. Therefore, here we restrict ourselves to explore only the MDM contribution.\\

We set the notation for the process as $ e^{+}(k_+) e^{-}(k_-) \to \pi^+(p_1) \pi^0(p_2) \pi^-(p_3) \pi^0(p_4)$, where in parenthesis are the corresponding 4-momenta. The total amplitude can be written as:
\begin{equation}
\mathcal{M}=\frac{e}{q^2} l^\mu J_\mu,
\end{equation}
where $1/q^2$ comes from the photon propagator ($q=k_+ +k_-$). The leptonic current $l^{\mu} \equiv  \bar{v}(k_+) \gamma^\mu u(k_-)$ is common to all the channels and  $J_\mu$ represents the four pion electromagnetic current. This last must fulfill the Bose-Einstein symmetry, by the interchange of the neutral pions, and $C$ invariance, by the interchange of the charged pions. These conditions have been enforced in previous studies in a generic way, by considering reduced amplitudes (not restricted by the symmetries) upon which the corresponding symmetries are applied. Then, the current is constructed by the combination of such reduced amplitudes \cite{Ecker2002, Czyz:2008kw}. Here, we follow a different handling of the amplitudes, which allows us to obtain explicit gauge invariant amplitudes, by a proper combination of different channels. 

The process has been studied, in the energy regime below 1 GeV, using effective models based on chiral symmetry and VMD \cite{Eidelman:1994zc,Ecker2002,Czyz:2008kw,Juran:2008kf}. Here, $\gamma^* \to 4\pi$ vertex is modelled in the VMD approach, by considering the relevant hadronic degrees of freedom in the energy range of our interest. 
The effective Lagrangian that is common to all the VMD based models \cite{Kroll:1967it,Bando:1984ej,Fujiwara:1984mp,Meissner:1987ge}, including the light mesons $\rho$, $\pi$ and $\omega$, in addition to the $\rho^\prime$ can be set as
\begin{eqnarray}
{\cal L}&=&
\sum_{V=\rho,\,\rho^\prime,\,\omega} \frac{e\, m_V^2}{g_V}\,V_\mu\, A^\mu +\sum_{V=\rho,\,\rho^\prime} g_{V\pi\pi}\,
\epsilon_{abc}\, V_\mu^a\, \pi^b \,\partial^\mu\, \pi^c  \nonumber \\
&+&\sum_{V=\rho,\,\rho^\prime} 
g_{\omega V\pi}\,\delta_{ab}\,\epsilon^{\mu\nu\lambda\sigma}\,\partial_\mu\, \omega_\nu\, \partial_\lambda\, V_\sigma^a\,  \pi^b \nonumber \\
&+&g_{3\pi}\,\epsilon_{abc}\,   \epsilon^{\mu\nu\lambda\sigma}\,\omega_\mu\, \partial_\nu\, \pi^a\,  \partial_\lambda\,  \pi^b\,  \partial_\sigma\, \pi^c .
\label{Lvmd}
\end{eqnarray}
$V$ and $A$ refer to the vector meson and  the photon fields respectively. The couplings are free parameters to be determined from experiment and the Latin indices denote the corresponding interacting fields. A contact term from the Wess-Zumino-Witten anomaly \cite{Wess:1971yu,Witten:1983tw} is included, with coupling constant $g_{3\pi}$.

\begin{figure}
\begin{center}
\includegraphics[scale=0.3]{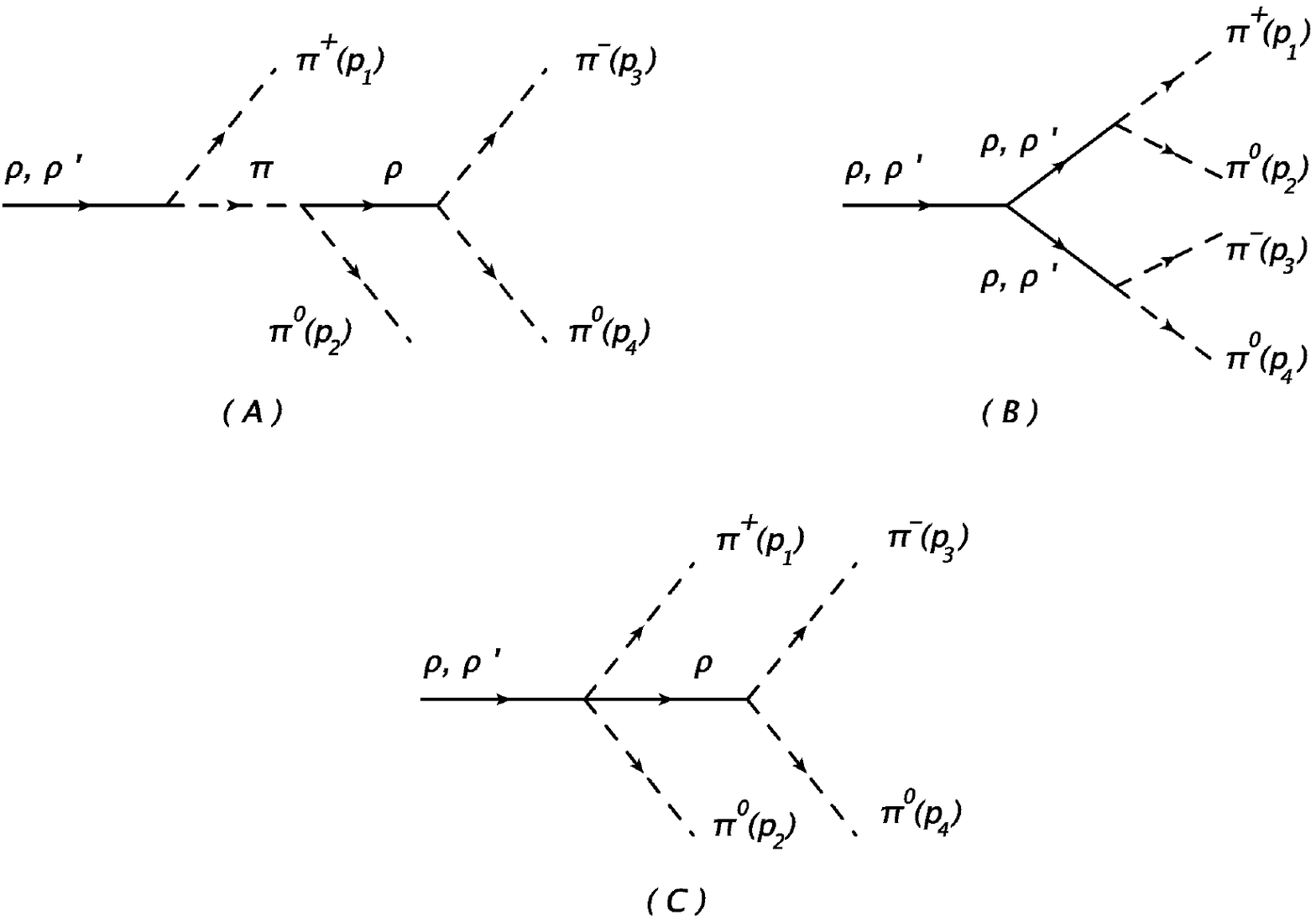}
\end{center}
\caption{Generic channels, linked by gauge invariance, relevant for the description of the $e^{+} e^{-} \to \pi^+ \pi^- 2 \pi^0$  process. The full set of diagrams are obtained by applying Bose-Einstein symmetry and Charge conjugation to each one.}
\label{procesos}
\end{figure}

\begin{figure}
\begin{center}
\includegraphics[scale=0.3, angle=-90]{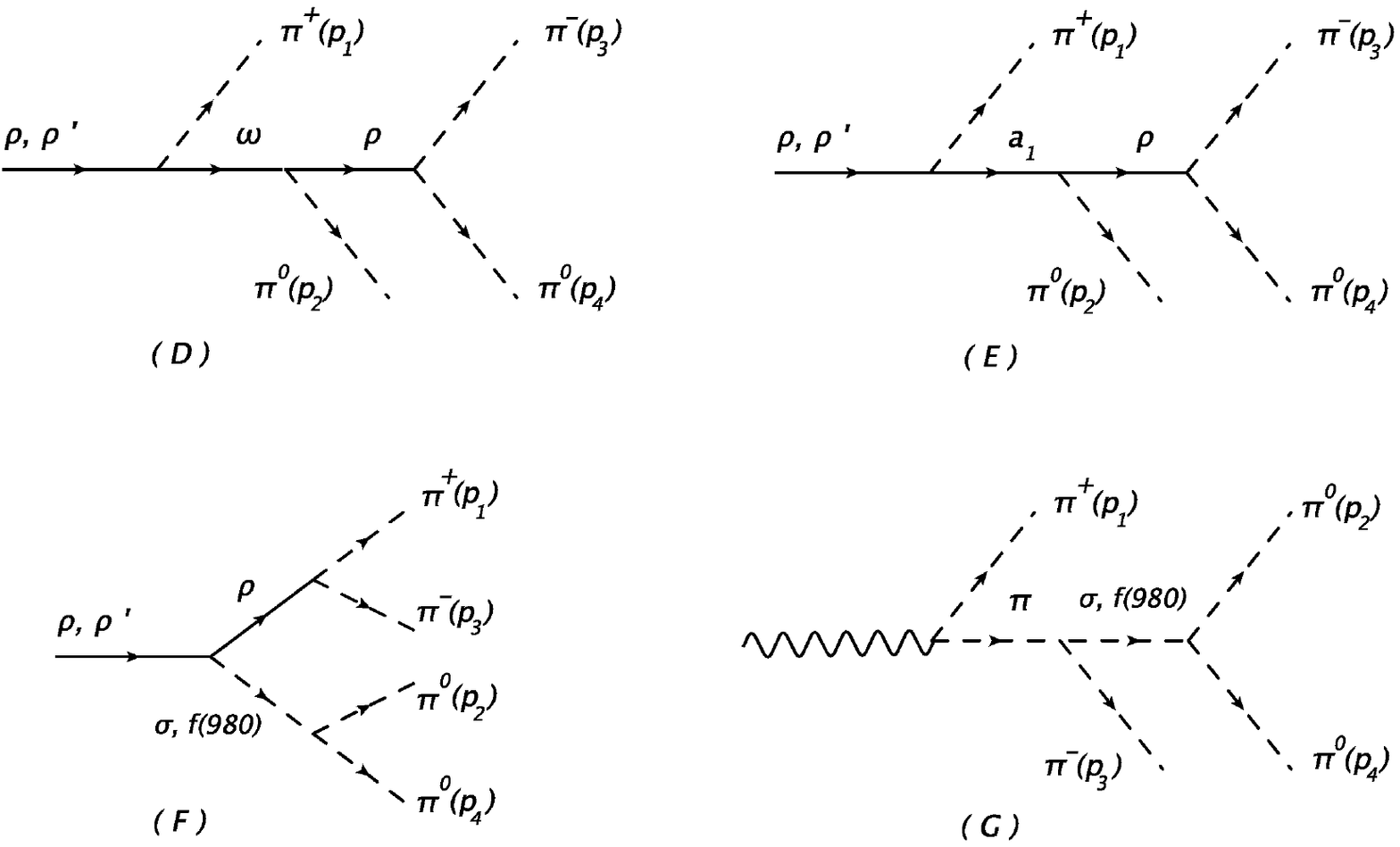}
\end{center}
\caption{Generic channels, involving the $\omega$, $a_1$ and scalar resonances, relevant for the description of the $e^{+} e^{-} \to \pi^+ \pi^- 2 \pi^0$ process. The full set of diagrams are obtained by applying Bose-Einstein symmetry and Charge conjugation to each one. They are gauge invariant by themselves.}
\label{procesos2}
\end{figure}

To model the four pion electromagnetic current, we consider the channels including the exchange of the $\pi$, $\omega$, $a_1$, $\sigma$, $f_0(980)$, $\rho$ and $\rho'(1450)$  mesons, as shown in Figures \ref{procesos} and \ref{procesos2}. Thus, we have seven generic channels (twenty four in total by including the neutral and charged pions exchange).
The contribution to the four pion electromagnetic current from each channel is denoted by $\mathcal{M}^{\mu}_{channel}(p_1,p_2,p_3,p_4)$. The additional amplitudes for the diagrams, corresponding to the neutral and charged pions exchange, will be denoted by just the momentum exchange in the functionality, with respect to the first case plotted in the Figures. Namely, the amplitude for the neutral pions exchange is denoted by $\mathcal{M}^{\mu}_{channel}(p_1,p_4,p_3,p_2)$, 
 for the charged pions exchange by
$\mathcal{M}^{\mu}_{channel}(p_3,p_2,p_1,p_4) $
and the application of both is denoted by
$\mathcal{M}^{\mu}_{channel}(p_3,p_4,p_1,p_2)$. Thus, the four pion electromagnetic current is made up of all these contributions
\begin{equation}
    J^\mu =\sum_{channel}  \mathcal{M}^{\mu}_{channel}.
\end{equation}
The energy range to be described goes from threshold up to 1.8 GeV. 
We proceed to discuss each channel in detail:\\

{\it Channels $A$, $B$ and $C$, linked by gauge invariance.}\\
In Figure \ref{procesos}, we show the diagrams where the intermediate states are the $\pi$ and $\rho$ mesons, labeled by ($A$), ($B$) and ($C$). These channels, with their corresponding neutral and charged pions exchange, are linked by the gauge invariance condition. This is an important observation since our description of channel ($B$), also called the $\rho$-channel, is not gauge invariant by itself. Here, we describe the three channels, and properly combine them to obtain an analytical gauge invariant amplitude. The details are presented in Appendix A.

The initial vector states, coupled to the photon from the leptonic current, are both the $\rho$ and the $\rho^\prime$ mesons. Here, we provide the explicit equations for the amplitude considering only the $\rho$ meson. The corresponding expression for the $\rho^\prime$ are similar, with the $\rho$ mass and couplings replaced by the ones of the $\rho^\prime$. A relative phase of $180^0$ is assumed, based on previous studies \cite{Ecker2002,Czyz:2008kw}.\\

{\it Channel A}\\
The amplitude for channel ($A$) in Figure \ref{procesos} is given by:
\begin{eqnarray}
     \mathcal{M}^{\mu}_{A}(p_1,p_2,p_3,p_4) = -e\,
 \frac{g_{\rho\pi\pi}^{3}}{g_{\rho}}\,m^{2}_{\rho}\,D^{\mu\alpha}_{\rho}[q]\,(q-2p_1)_{\alpha}\nonumber\\
 S_{\pi}[q-p_1](q-p_1+p_2)_{\gamma}\,D^{\gamma\eta}_{\rho^{-}}[s_{43}]\,r_{43\eta},
 \label{MA1}
\end{eqnarray}
where $s_{ij}\equiv p_i + p_j$, $ r_{ij} \equiv  p_i-p_j$. $S_{\pi}[q]\equiv i/(q^{2}-m^{2}_{\pi}) $ is the pion propagator with mass $m_{\pi}$. The propagator of vector mesons with mass $m_V$ and decay width $\Gamma_{V}$ is assumed to have the complex mass form. This is consistent with the inclusion of the absorptive corrections and gauge invariance \cite{Baur:1995aa,Argyres:1995ym,Beuthe:1996fe,LopezCastro:1999xg} 
\begin{equation}
    D^{\alpha\mu}_V[p]=-iD_V[p] \left(g^{\alpha\mu}-\frac{p^{\alpha}p^{\mu}}{m^{2}_V-i\,m_V\Gamma_V}\right),
\end{equation} 
where $D_V[p]\equiv 1/(p^{2}-m^{2}_V + i\,m_V\Gamma_V)$. We work in the isospin limit. However, in some cases we label the vector mesons including their corresponding charges, to easy the reading from the corresponding Feynman diagrams.
The energy dependence of the width will be used only for the $\rho$ meson  
\begin{equation}
\Gamma_\rho(s)=\Gamma_\rho \left(\frac{m_\rho}{\sqrt{s}}\right)^5\,\left[\frac{\lambda(s,m_\pi^2,m_\pi^2)}
{\lambda(m_\rho^2,m_\pi^2,m_\pi^2)}\right]^{3/2},
\end{equation}
where $\lambda(x,y,z)$ is the K\"allen function and $\Gamma_\rho$ is the constant decay width. For the $\rho^\prime$ the difference between considering a constant width and an energy dependent one (considered to be dominated by the $\pi\pi$ channel) is within the uncertainties of the corresponding parameters. 
In the isospin symmetry limit the amplitude gets the simplified form:
\begin{equation}
 \begin{split}
 \mathcal{M}^{\mu}_{A}(p_1,p_2,p_3,p_4)&=i\,e\,C
 D_{\rho^{-}}[s_{43}]\,r^{\gamma}_{43}\,\,z_{12\gamma}\,\frac{x_1{}^{\mu}}{x_1\cdot q},\\
 \end{split}
 \label{MAr}
\end{equation}
where $x_1 \equiv q-2\,p_1$, $z_{12} \equiv q-p_1+p_2$ and $C\equiv (g_{\rho\pi\pi}^{3}/g_{\rho})\,m^{2}_{\rho}\,D_{\rho}[q] $. A similar expression corresponds to the case for the $\rho^\prime$, with $C \to (g_{\rho\pi\pi}^{2}g_{\rho^\prime\pi\pi}/g_{\rho^\prime})\,m^{2}_{\rho^\prime}\,D_{\rho^{\prime}}[q]$.
In the previous analysis \cite{GarciaGudino:2013alv,David2015}, given the scarce information on the $\rho^\prime$ decay modes, the following combination of coupling constants for the $\rho^\prime$ was assumed:
\begin{equation}
\frac{m^2_{\rho '}g_{\rho'}}{g_{\rho ' \pi \pi}} = X \frac{m^2_{\rho}g_{\rho} }{g_{\rho \pi \pi}},
 \label{capprox}
\end{equation}
where the proportionality constant $X=1$ was taken. The idea behind this assumption was to resemble typical VMD relations, expecting the particularities of the radial excitation to drop out when considering the ratios. Deviations to this relation of up to 20\% were explored and considered in the final result uncertainty. 
In this work, we do not rely on this ansatz, but make use of additional theoretical and experimental information. In Table \ref{fitall}, we show the values of the couplings, consistent with a large set of observables \cite{Avalos:2022txl}, which favors $X= 1.3\pm 0.4$. This is a larger value, but in agreement with the previous one within uncertainties.\\

{\it Channel B}\\
In Figure \ref{procesos} ($B$), we show the so-called $\rho$-channel, which includes the $\rho\rho\gamma$ vertex, relevant for the extraction of the MDM of the $\rho$ meson. 
The photon from the leptonic side, couples to the vector meson, which then couples to a triple vector vertex. This defines the $\rho$ electromagnetic vertex, times a global constant $g_{\rho \pi \pi }$, accounting for the strong process. 
For the $\rho^\prime$ triple vector meson vertex, we take the same structure as for the $\rho$ case, this assumption has been found to be appealing \cite{Czyz:2008kw}, with the corresponding difference in couplings. 
The amplitude is given by:
\begin{align}
\mathcal{M}^{\mu}_{B}(p_1,p_2,p_3,p_4) =& 
-e\,\frac{g_{\rho\pi\pi}^{3}}{g_{\rho}}\,m^{2}_{\rho}\,D^{\mu\alpha}_{\rho}[q]\,r_{12\gamma}\nonumber\\
      &\,D^{\gamma\lambda}_{\rho^+}[s_{21}]\,\Gamma^{1}_{\alpha\lambda\delta}\,D^{\delta\eta}_{\rho^-}[s_{43}]\,r_{43\eta},
      \label{MB1}
\end{align}
where $\Gamma^{1}_{\alpha\lambda\delta} \equiv (1+i\,\gamma)\,\Gamma_{\alpha\lambda\delta}$, is the absorptive corrected vertex at one-loop, consistent with gauge invariance \cite{Baur:1995aa,Argyres:1995ym,Beuthe:1996fe,LopezCastro:1999xg}, and $\gamma\equiv \Gamma_V/M_V$. The tree-level vertex Eq.~(\ref{vertex}) for this momentum configuration takes the form:
\begin{flalign}
      &\Gamma_{\alpha\lambda\delta} = \\
      &g_{\lambda\delta}\,\,Q_{1}{}_{\alpha} + \beta_0\,(q_{\delta}\,\,g_{\alpha\lambda}-q_{\lambda}\,\,g_{\delta\alpha}) + s_{21\lambda}\,\,g_{\delta\alpha} - s_{43\delta}\,\,g_{\alpha\lambda},\nonumber
\label{G0vertex}
\end{flalign}
where $q = s_{21} + s_{43}$ and $Q_{1} \equiv s_{43}-s_{21}$. We have set $\beta_0\equiv \beta(0)$ for simplicity, the dependence on $q^2$ is accounted by the neutral vector mesons coupled to the photon.
The simplified amplitude is:
 \begin{equation}
 \begin{aligned}
 \mathcal{M}^{\mu}_{B}&(p_1,p_2,p_3,p_4)=\\
 &i\,e\,C\,\Big\{\Big(D_{\rho^{-}}[s_{43}]-D_{\rho^{+}}[s_{21}]\Big)\,\frac{r_{43}\cdot r_{12}}{Q_{1}\cdot q}Q^{\mu}_{1} \\
 & +(1+i\,\gamma)\,D_{\rho^{-}}[s_{43}]\,D_{\rho^{+}}[s_{21}]\\ 
 &\beta_0\,\big(q\cdot r_{12}\,r^{\mu}_{43}-q\cdot r_{43}\,r^{\mu}_{12}\big)\Big\}.\label{MBr}
 \end{aligned}
\end{equation}
 
{\it Channel C}\\
In Figure \ref{procesos} ($C$), we show the process driven by a contact term ($\rho\rho\pi\pi$), whose amplitude can be written in a general form as: 
\begin{eqnarray}
     \mathcal{M}^{\mu}_{C}(p_1,p_2,p_3,p_4)=i\,e\, \frac{g_{\rho\pi\pi}\,g_{\rho\rho\pi\pi}}{g_{\rho}}m^{2}_{\rho}\nonumber\\
     D^{\mu\alpha}_{\rho^{0}}[q]\,\Gamma^{1}_{\alpha\gamma}\,D_{\rho^{-}}[s_{43}]\,r^{\gamma}_{43}.\label{MCr}
\end{eqnarray}
The effective coupling $g_{\rho\rho\pi\pi}$ and the general vertex $\Gamma^{1}_{\alpha\gamma}$ are fixed by requiring gauge invariance of the sum of the ($A$), ($B$) and ($C$) amplitudes. To obtain the gauge invariant amplitude associated to the combination of these channels, we profit from the Ward-Takahashi identity, fulfilled by the $VV\gamma$ vertex. Instead of leaving the counter-term as a general requirement, as done before. This allows to trace back the origin of the different contributions that combine to build the gauge invariant amplitude.
In order to get the gauge invariant amplitude, we use the combination of the following amplitudes:
\begin{eqnarray}
\mathcal{M}^{\mu}_{ABC_{24}} &=& \mathcal{M}^{\mu}_{A}(p_1,p_2,p_3,p_4)
+\mathcal{M}^{\mu}_{A}(p_3,p_4,p_1,p_2)\nonumber\\
&&+\mathcal{M}^{\mu}_{B}(p_1,p_2,p_3,p_4)\\
&&+\mathcal{M}^{\mu}_{C}(p_1,p_2,p_3,p_4)+\mathcal{M}^{\mu}_{C}(p_3,p_4,p_1,p_2),\nonumber
\end{eqnarray}
where we have set the notation of the amplitude to refer to the three channels and the neutral pions momenta.
The gauge invariant amplitude is then:
\begin{flalign}
 &\mathcal{M}^{\mu}_{ABC_{24}} =  i\,e\, C \,\Big\{L^{\mu}(x_1,x_3) \nonumber\\
 & \Big(D_{\rho^{-}}[s_{43}]\,r_{43}\cdot z_{12} - D_{\rho^{+}}[s_{21}]\,r_{12}\cdot z_{34}\Big)\nonumber\\
& +r_{43}\cdot r_{12}\Big(D_{\rho^{-}}[s_{43}]\,L^{\mu}(Q_{1},x_3)-D_{\rho^{+}}[s_{21}]\,L^{\mu}(Q_{1},x_1)\Big)\nonumber\\
& +(1+i\,\gamma)\,D_{\rho^{-}}[s_{43}]\,D_{\rho^{+}}[s_{21}]\nonumber\\
&\beta_0\,\Big(q\cdot r_{12}\,r^{\mu}_{43}-q\cdot r_{43}\,r^{\mu}_{12}\Big)\Big\},
\label{abc24}
 \end{flalign}
where $x_1 \equiv q-2\,p_1$, $x_3 \equiv q-2\,p_3$ and we defined the gauge invariant tensor:
\begin{equation}
    L^{\mu}(a,b) \equiv \frac{a^{\mu}}{a\cdot q} - \frac{b^{\mu}}{b\cdot q}.\label{Lmunu}
\end{equation}
A similar expression is obtained by adding the remaining amplitudes:
\begin{eqnarray}
 \mathcal{M}^{\mu}_{ABC_{42}} &=& \mathcal{M}^{\mu}_{A}(p_1,p_4,p_3,p_2)+\mathcal{M}^{\mu}_{A}(p_3,p_2,p_1,p_4)\nonumber\\
 &&+\mathcal{M}^{\mu}_{B}(p_1,p_4,p_3,p_2)\label{abc42}\\
 &&+\mathcal{M}^{\mu}_{C}(p_1,p_4,p_3,p_2)+\mathcal{M}^{\mu}_{C}(p_3,p_2,p_1,p_4),\nonumber
\end{eqnarray}
which in practice corresponds to the $p_2 \leftrightarrow p_4$ exchange. 
The charged pions exchange was already used to build the gauge invariant structures. The details to obtain the gauge invariant amplitude are described in Appendix A.\\

\begin{figure}[t]
\begin{center}
\includegraphics[scale=0.67]{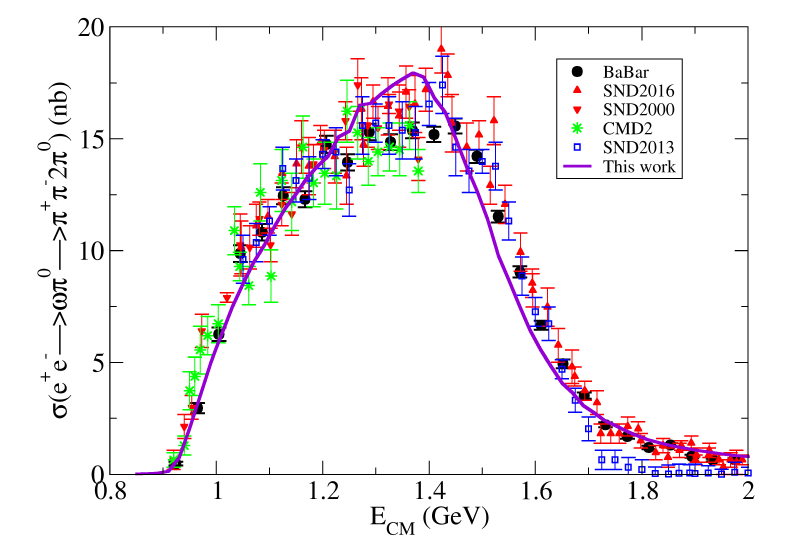}
\end{center}
\caption{$e^{+}e^{-}\rightarrow\omega\pi^{0}\rightarrow\pi^{+}\pi^{-}2\pi^{0}$ cross section, the solid line is the result obtained in this work using the parameters given in Table \ref{fitall}data and the symbols correspond to the data from BaBar  \cite{BaBar:2017zmc} and the rescaled data from SND \cite{ACHASOV200029,Achasov:2003bv,Achasov:2009zz} and CMD2 \cite{CMD-2:2003bgh}) measured for the final state $\omega\rightarrow \pi^{+}\pi^{-}\pi^{0}$.}
\label{FitD}
\end{figure}

{\it Channel D}\\
In Figure \ref{procesos2} ($D$), we show the channel corresponding to the contribution of the $\omega$. This decay to three pions, via the $\rho$ and $\rho^\prime$ mesons intermediate states and a contact term. The intermediate states are threefold, as they can be charged and neutral. The amplitude can be set as:
\begin{eqnarray}
    \mathcal{M}^{\mu}_{D}(p_1,p_2,p_3,p_4) &=&-i\,e\,\Big(C_d + e^{i\theta}\,C^{\prime}_d\Big)\,D_{\omega}[q-p_2] \nonumber\\
    &&\mathcal{A}[(q-p_2)^{2}]\,V^{1\mu}_{\omega},
\label{Dr1}
\end{eqnarray}
where
\begin{flalign}
&\mathcal{A}[(q-p_2)^{2}] = 6\,g_{3\pi}  \nonumber\\ &+2\,g_{\rho\pi\pi}\,g_{\omega\rho\pi}\Big(D_{\rho^{0}}[s_{13}]+D_{\rho^{+}}[s_{41}]+D_{\rho^{-}}[s_{43}]\Big)\nonumber\\
&+ 2\,g_{\rho^{\prime}\pi\pi}\,g_{\omega\rho^{\prime}\pi}\Big(D_{\rho^{\prime}}[s_{13}]+D_{\rho^{\prime}}[s_{41}]+D_{\rho^{\prime}}[s_{43}]\Big),    
\end{flalign}
and
\begin{equation}
    V^{1\mu}_{\omega} \equiv \epsilon_{\alpha\eta\beta\sigma}\,\epsilon^{\mu\gamma\chi\sigma}\,q_{\gamma}\,{p_2}_{\chi}\, p^{\alpha}_{1}\,p^{\eta}_{3}\,p^{\beta}_{4}.
\end{equation}
The $\rho$ and $\rho^\prime$ are added with a relative phase $\theta$ and the corresponding coefficients are
\begin{equation}
    C_d \equiv \frac{g_{\omega\rho\pi}}{g_{\rho}}\,m^{2}_{\rho}\,D_{\rho}[q],\qquad C^{\prime}_d \equiv \frac{g_{\omega\rho^{\prime}\pi}}{g_{\rho^{\prime}}}\,m^{2}_{\rho^{\prime}}\,D_{\rho^{\prime}}[q].
\end{equation}
A similar expression, corresponding to the neutral pions exchange ($p_2\leftrightarrow p_4$) completes the amplitudes for this channel. They are gauge invariant by themselves.\\
In the previous analysis \cite{Gudio2012,GarciaGudino:2013alv,David2015}, we determined the $g_{\rho^{\prime\omega\pi}}$ coupling by a fit to the SND \cite{Achasov:2003bv,Achasov:2009zz} and preliminary BaBar data \cite{druzhinin2007,Druzhinin:2011qd} for this channel, with a  relative phase $\theta$ between the $\rho$ and $\rho^{\prime}$ of 180$^{0}$. The error bar on the coupling accounted for the difference to fit both data sets individually.
Here, we have redone the analysis of Ref. \cite{Avalos:2022txl}, extending the observables to include the $\omega$ channel data from BaBar \cite{BaBar:2017zmc}. Previously, it was considered as a prediction. In Table \ref{fitall}, we show the parameters obtained. They are in agreement with the previous ones \cite{Avalos:2022txl}, but improve on the uncertainties of the parameters associated to the $\rho^\prime$.\\
In Figure \ref{FitD}, we show the $\omega$ channel cross section obtained using the parameters given in Table \ref{fitall} and the experimental data reported by BaBar \cite{BaBar:2017zmc}. We also plot the SND \cite{ACHASOV200029,Achasov:2013btb,Achasov:2016zvn} and CMD2\cite{CMD-2:2003bgh} data, obtained from rescaling the measured process $e^{+}e^{-}\to \pi\pi\gamma$. We observe that the theoretical result from this work makes a good description of them. Thus, this intermediate energy region is well under control, setting the baseline for the MDM determination, as we show below.\\
 \begin{table} 
    \centering 
    \begin{tabular}{|c|c|}
    \hline
    \textbf{Parameter} &  \textbf{Value} \\
    \hline 
    $g_{\rho \pi \pi }$ & 5.95    $\pm$   0.08 \\
    $g_{\rho}$  &  4.96    $\pm$  0.09\\
    $g_{\omega}$  &  16.62   $\pm$    0.47 \\
    $g_{\omega\rho\pi}$ (GeV$^{-1})$ & 11.29  $\pm$ 0.38 \\ 
    $g_{\rho^\prime \pi \pi }$ &  5.8    $\pm$    0.44\\  
    $g_{\omega\rho^\prime\pi}$ (GeV$^{-1})$ & 3.61  $\pm$ 0.74 \\ 
    $g_{3 \pi }$ (GeV$^{-3}$)& -53.49 $\pm$   7.19\\
    $g_{\rho^\prime}$  &   12.85      $\pm$  0.4  \\ 
    $\theta$ (in $\pi$ units)&  0.9    $\pm$   0.04\\  
    \hline
    \end{tabular}
\caption{Parameters obtained from a fit to a set of observables as in Ref. \cite{Avalos:2022txl} and adding the $e^+e^-\to\omega\pi\to\pi^+\pi^-2\pi^0$ channel data from BaBar.}
\label{fitall}
\end{table}

{\it Channels E, F and G}\\
These channels involve, in the intermediate state,  the $a_1$ axial vector meson, and the $\sigma$(600) and $f_0$(980) scalar particles, as depicted in Figure \ref{procesos2}. Here, we present the corresponding amplitudes in explicit gauge invariant form.
These channels are suppressed in the whole region of study,
 as it has been shown in the previous analysis \cite{GarciaGudino:2013alv,David2015}. They are included to compare the model near threshold, where they become relevant, with data presented by BaBar \cite{BaBar:2017zmc} (well below the energy region of interest for the $\rho$ MDM), and to exhibit the dependence on the large uncertainties associated to the parameters of these channels.\\ 

Channel $E$ involves the axial vector meson $a_1$. The effective lagrangian describing the interaction between the  $a_1(q)-\rho(k)-\pi(p)$, where in parenthesis is the corresponding momentum, is given by \cite{Isgur:1988vm}:
\begin{equation}
    \mathcal{L}_{a_1} = 2\,g_{a_1\rho\pi} \Big(\rho_{\mu}\,a^{\mu}_1\,q\cdot k-\partial_{\nu}\,\rho^{\mu}\,\partial_{\mu}\,a^{\nu}_1\Big).
\end{equation}

The amplitude, after simplifications, becomes
\begin{eqnarray}
     \mathcal{M}^{\mu}_{E}(p_1,p_2,p_3,p_4) &=&-i\,e\,C_a\,D_{\rho^{-}}[s_{43}]\,D_{a_1}[q-p_1]\\
     & &F^{\mu\alpha}(q-p_1,q)\,F_{\alpha\beta}(q-p_1,s_{43})\,r^{\beta}_{43},\nonumber
\end{eqnarray}
where $C_a \equiv (g_{a_1\rho\pi}^{2}\,g_{\rho\pi\pi}/g_{\rho})m^{2}_{\rho}\,D_{\rho}[q]$ and the tensor function is given by:
\begin{equation}
\begin{aligned}
 F_{\mu\alpha}(a,b) \equiv a\cdot b\,\,g_{\mu\alpha}-a_{\mu}b_{\alpha}.
 \label{LamOme1}
\end{aligned}
\end{equation}
The gauge invariance is explicit from the structure of this tensor.
 Similar expressions are obtained for the corresponding charged and neutral pions exchange. In all cases the $\rho^\prime$ contribution is added with the same considerations as in the previous channels. The coupling $g_{a_1\rho\pi}= 3.25 \pm 0.3$ GeV$^{-1}$ is determined from the $a_1 \to \rho\pi $ decay width, and the $a_1$ total decay width is considered as a constant. Due to the lack of experimental information, the coupling for the $\rho^\prime$ is considered to be the same as for the $\rho$.\\

{\it Channel F}\\
This channel involves the $\rho$ meson and the scalar particle intermediate states $\sigma$(600) and $f_0$(980). 
The effective Lagrangian for this channel is given by:
\begin{equation}
    \mathcal{L}_S = g_{V_1V_2S}\,g_{\mu\nu}\,V^{\mu}_{1}\,V^{\nu}_{2}\,S + g_{SP_1P_2}\,S\,P_1\,P_2,
\end{equation}
where $g_{V_1V_2 S}$ and $g_{S P_1 P_2}$ are the effective couplings.
The simplified amplitude is given by
\begin{eqnarray}
\mathcal{M}^{\mu}_{F_{\sigma}}(p_1,p_2,p_3,p_4) &=&i\,e\,C_{\sigma}\,D_{\sigma}[s_{24}]\,D_{\rho^{0}}[s_{31}]\nonumber\\
&&F^{\mu\beta}(s_{31},q)\,{r_{31}}_{\beta},   
\end{eqnarray}
where the propagator of the $\sigma$ meson is $D_{\sigma}[s]\equiv i/(s-m_\sigma^2+i m_\sigma \Gamma_\sigma)$ and $C_{\sigma} \equiv (g_{\sigma\pi\pi}\,g_{\rho\rho\sigma}\,g_{\rho\pi\pi})/g_{\rho})\,m^2_{\rho}\,D_{\rho}[q]$.

\noindent Notice that this channel has neither neutral nor charged pions exchange, since they are emitted from the same vertex. 
Using VMD relations, the coupling $g_{\rho\rho\sigma} = -(e/g_{\rho})\,g_{\rho\sigma\gamma}$ is determined using the $\rho\to\sigma\gamma$ decay width, which gives $g_{\rho\sigma\gamma}=0.63 \pm 0.15$ and the coupling $g_{\sigma\pi\pi} =3.7 \pm 1.6 $ is determined from the $\sigma\to \pi\pi$ decay width.
The contribution from the $f_0(980)$ is similar, with the mass and width replaced accordingly, and we assume the couplings to be the same as for the $\sigma$.\\

{\it Channel G}\\
This is a channel where neither $\rho$ nor $\rho^\prime$ do resonate. Thus, we have a direct coupling of the photon to two pions. It includes an intermediate scalar particle, that can be both the $\sigma$(600) and $f_0(980)$. 
The neutral pions emission comes from a single vertex. Thus, we end up having only two diagrams due to the charged pions exchange. The sum of the corresponding amplitudes is given by:
\begin{equation}
 \mathcal{M}^{\mu}_{G} = i\,e\,(g_{\sigma\pi\pi})^{2}\,D_{\sigma}[s_{42}]\,L^{\mu}(x_1,x_3),
\end{equation}
which is gauge invariant. The contribution from the $f_0(980)$ is similar, with the mass and width replaced accordingly, and we assume the couplings to be the same as for the $\sigma$. As we pointed out above, the information on the couplings and parameters of the scalars are not well determined, being a strong source of uncertainties in the low energy regime.

\begin{figure}[t]
\begin{center}
\includegraphics[scale=0.52]{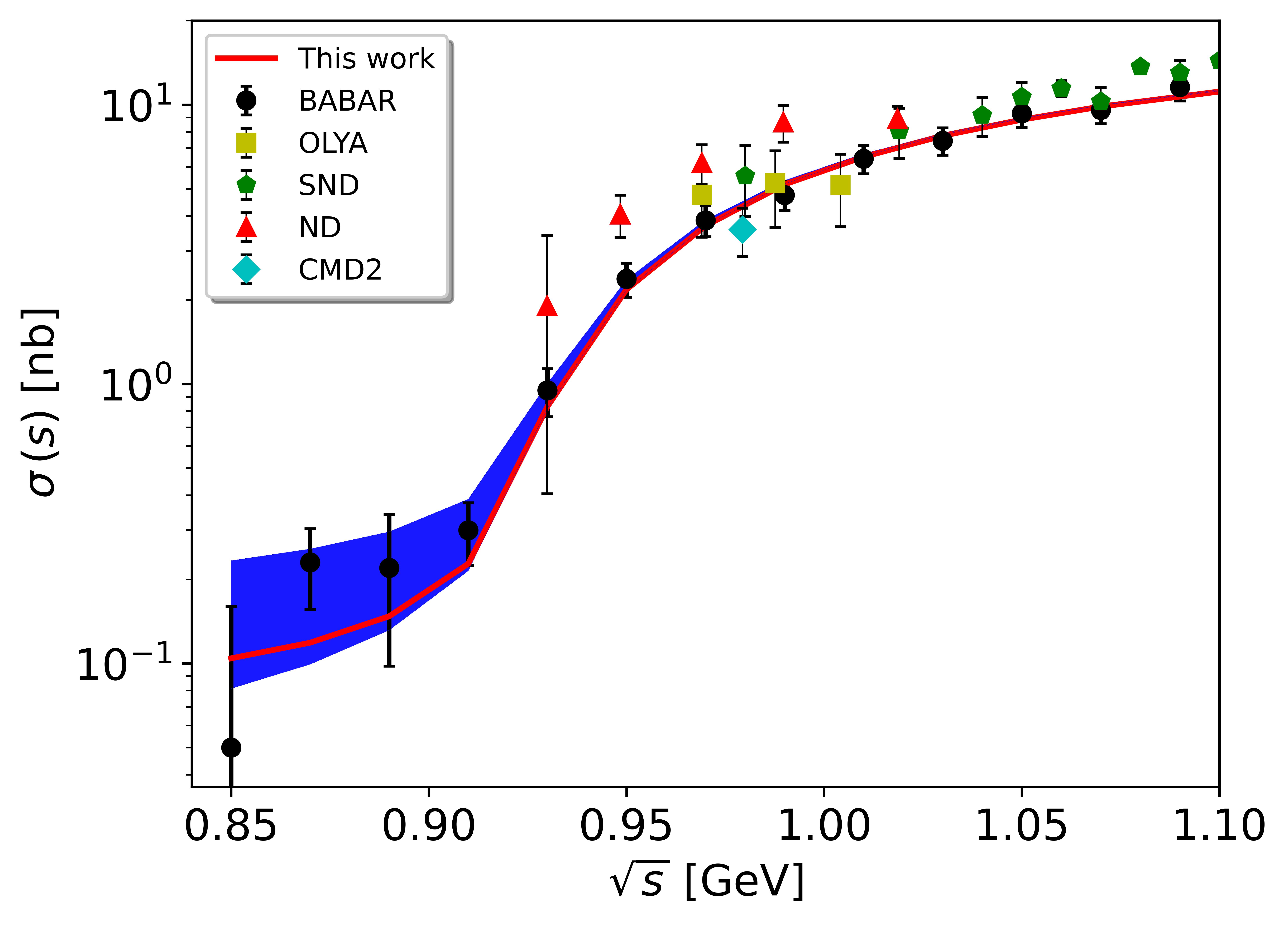}
\end{center}
\caption{Total cross section  $e^+ e^- \to \pi^+ \pi^- 2 \pi^0$ in the energy region below 1.1 GeV, compared with several experimental data: SND, BaBar, OLYA, CMD2 and ND.}
\label{lowcross}
\end{figure}	

In Figure \ref{lowcross}, we plot the cross section in the low energy region (below 1.1 GeV). 
The result from our model (solid line) is compared with experimental measurements from SND \cite{Achasov:2003bv,Achasov:2009zz},  BaBar \cite{BaBar:2017zmc}, OLYA, CMD2 and ND \cite{Cosme:1978qe,Bacci:1980zs,Kurdadze:1986tc,CMD-2:1998gab} (symbols), which are properly described. In this region, the cross section is dominated by the ($D$) and ($G$) channels, consistent with what has been found in previous analysis \cite{Czyz:2008kw}. The uncertainties associated to these channels come mainly from the scalar particles parameters, and the interference among the channels. They are displayed in Figure \ref{lowcross} by the blue shaded band. Notice that the band fades out as energy increases, and therefore have no impact in the analysis for the determination of the $\rho$ MDM.
Also note that the theoretical result quoted in Ref.\cite{BaBar:2017zmc}, for the low energy region, corresponds to a single contribution while our description involves several channels.

\section{The $\rho$ MDM from the total cross section}	
We have computed the total cross section following the PDG convention \cite{pdg}, and neglected the electron mass:
\begin{equation}
    d\sigma = \frac{(2\pi)^{4}\,\overline{|\mathcal{M}|^{2}}}{4\,\sqrt{(k_+\cdot k_-)^{2}}}\,\delta^{4}\Big(q-\sum^{n}_{i=1}p_i\Big)\prod^{n}_{i=1}\frac{d^{3}p_i}{(2\pi)^{3}\,2E_i}.\label{sigmaee4pi2}
\end{equation}
The averaged squared amplitude  $|\overline{\mathcal{M}|^{2}}$ is built up by the leptonic and hadronic parts discussed above: 
\begin{equation}
    \overline{|\mathcal{M}|^{2}} = \frac{e^2}{q^{4}}\,l_{\mu\nu}\,h^{\mu\nu}, 
\label{BarM}
\end{equation}
where the $1/q^{4}$ factor comes from the photon propagator,  $h^{\mu\nu} \equiv J^{\mu}\,J^{\nu\dagger}$ is the hadronic tensor obtained from the hadronic currents discussed above, and $l_{\mu\nu}$ is the leptonic tensor, given by:
\begin{equation}
    l_{\mu\nu} = k_{+\mu}\,k_{-\nu}+k_{-\mu}\,k_{+\nu}-\frac{q^{2}}{2}\,g_{\mu\nu}.
\label{lmunu2}
\end{equation}
The kinematical variables and limits are chosen following Ref.\cite{Kumar:1969jjy}. The integration is performed numerically, using a Fortran code and the Vegas subroutine \cite{PETER1978192}.

\begin{figure}[t]
\begin{center}
\includegraphics[scale=0.6]{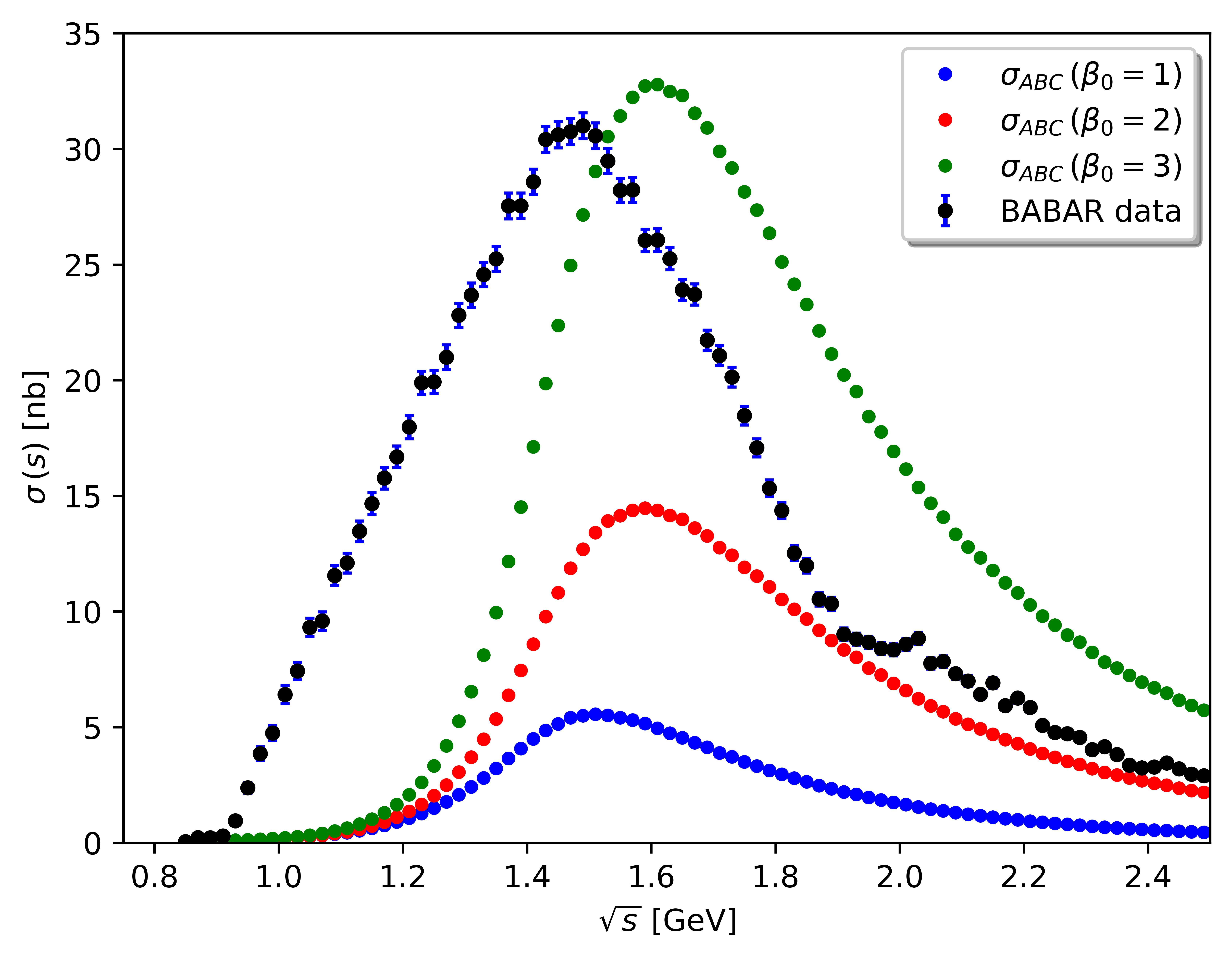}
\end{center}
\caption{BaBar data for the $e^{+}\,e^{-}\to 2\pi^{0}\pi^{+}\pi^{-}$ total cross section (symbols with error bars) and the contribution from $A$, $B$ and $C$ channels for $\beta_0=1,\ 2 ,\ 3$ (dotted lines).}
\label{betasense}
\end{figure}

In order to exhibit the dependence on the $\beta_0$ parameter, in Figure \ref{betasense}, we plot the cross section from $A$, $B$ and $C$ channels amplitude. We consider $\beta_0 =$1, 2 and 3. These are compared with the experimental data from BaBar. The plots show that the main sensitivity to $\beta_0$ is in the region from 1.4 GeV to 1.8 GeV. Thus, we consider the region up to 1.8 GeV to determine the $\rho$ MDM. For energies above 1.8 GeV there are structures that are not captured in the current description.

\begin{figure}[t]
\begin{center}
\includegraphics[scale=0.55]{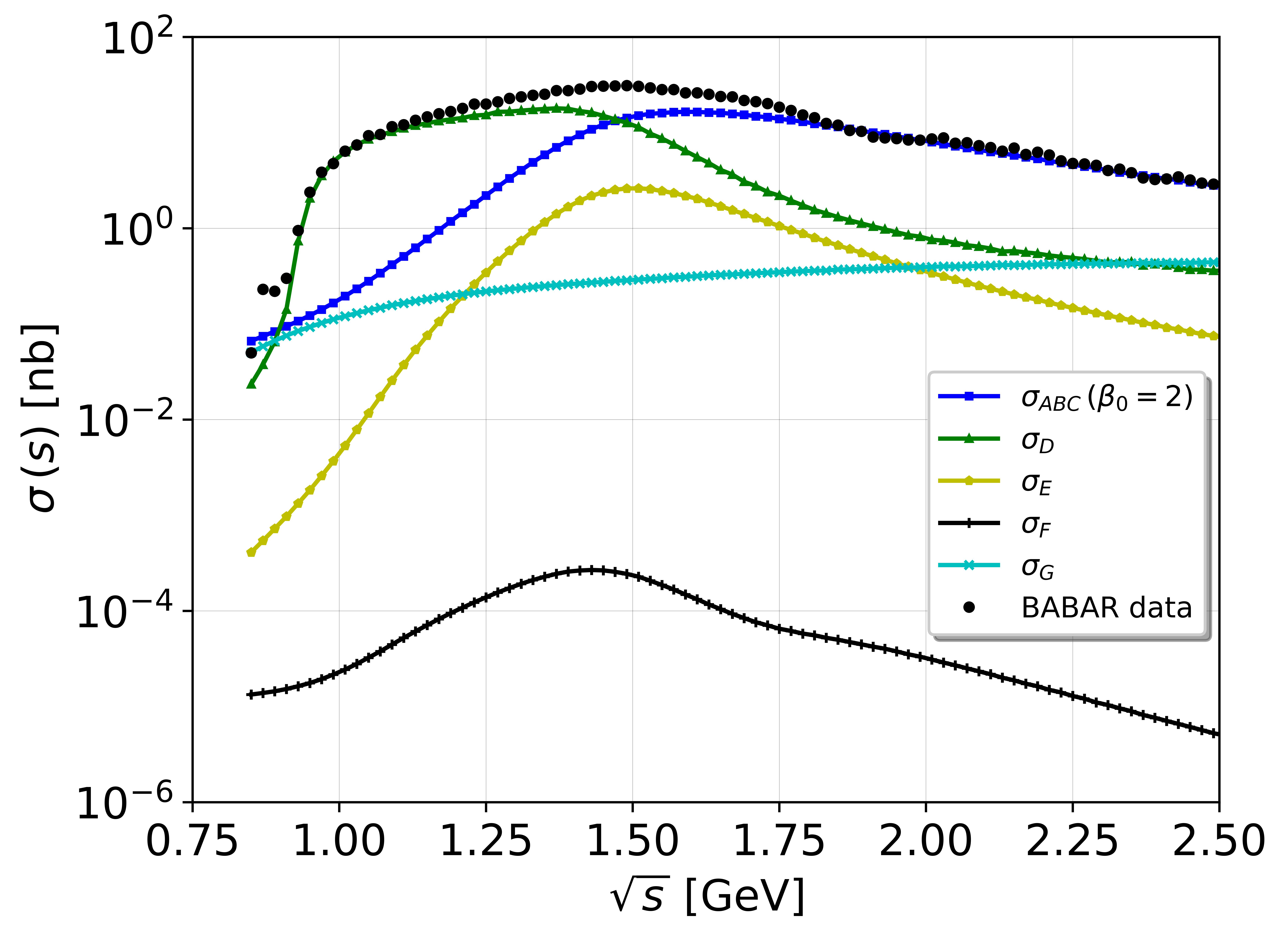}
\end{center}
\caption{BaBar data for the $e^{+}\,e^{-}\to 2\pi^{0}\pi^{+}\pi^{-}$ total cross section (circle symbols) and the contributions from all the channels.
The contribution from $A$, $B$ and $C$ channels is set to $\beta_0=2$.}
\label{allchannels}
\end{figure}

In Figure \ref{allchannels}, we plot the total cross section data from BaBar \cite{BaBar:2017zmc} and the contribution from each channel ($A$, $B$ and $C$ channels are together). They include the amplitudes for $\rho$ and $\rho^\prime$ and their corresponding interference. The interference among different channels are not shown, but accounted for in the analysis. Error bars are not displayed for the sake of clarity. The energy region below 1.4 GeV is dominated by the $\omega$ channel ($D$) and the interference with the other channels. Channels $F$ and $G$ are largely suppressed, they become less suppressed near threshold. The corresponding large uncertainties play a role, mainly through the interference with the $\omega$ channel. This is well below the region where $\beta_0$ is important. Above 1.4 GeV the contribution from $ABC$ channels surpass the $\omega$ contribution, becoming the leading one. In that region, channel $E$ is larger than channels $F$ and $G$, but still below the $ABC$ and $D$ channels. A fit to the BaBar\cite{BaBar:2017zmc} total cross section data, considering $\beta_0$ as the only free parameter, favors $\beta_0= 2.05 \pm 0.07$ with a $\chi^2/dof= 1.3$. In order to properly obtain the static limit value, we look to the electric charge form factor
\begin{flalign}
    &|F_{\rho }\left(0\right)| =\\
&\lim_{q^2 \to 0} 
\left\lvert \frac{g_{\rho  \pi \pi }m_{\rho }^2 }{g_{\rho}}D_\rho[q^2]-
\frac{g_{\rho^\prime  \pi \pi }m_{\rho^\prime }^2 }{g_{\rho^\prime}}D_{\rho^\prime}[q^2]
\right\rvert=1.\nonumber
\label{fformarec}
\end{flalign}
Using the value of the parameters as given in Table \ref{fitall}, we obtain the value for the left hand side limit $|F_{\rho }\left(0\right)|=0.75 \pm 0.05$, where the error comes mainly from the $\rho^\prime$ parameters. We normalize $\beta_0$ to this value to have the MDM properly defined. Thus, the value of the $\rho$ meson MDM from the BaBar data is  
\begin{equation}
\mu_\rho = 2.7 \pm 0.3 \ \ \text{in} \ (e/2 m_\rho) \ \text {units}.
\label{resfinbeta}
\end{equation}
The error bar considers the uncertainties from the fit and the one from the charge normalization, this last being the dominant. 

\begin{figure}[t]
\begin{center}
\includegraphics[scale=0.57]{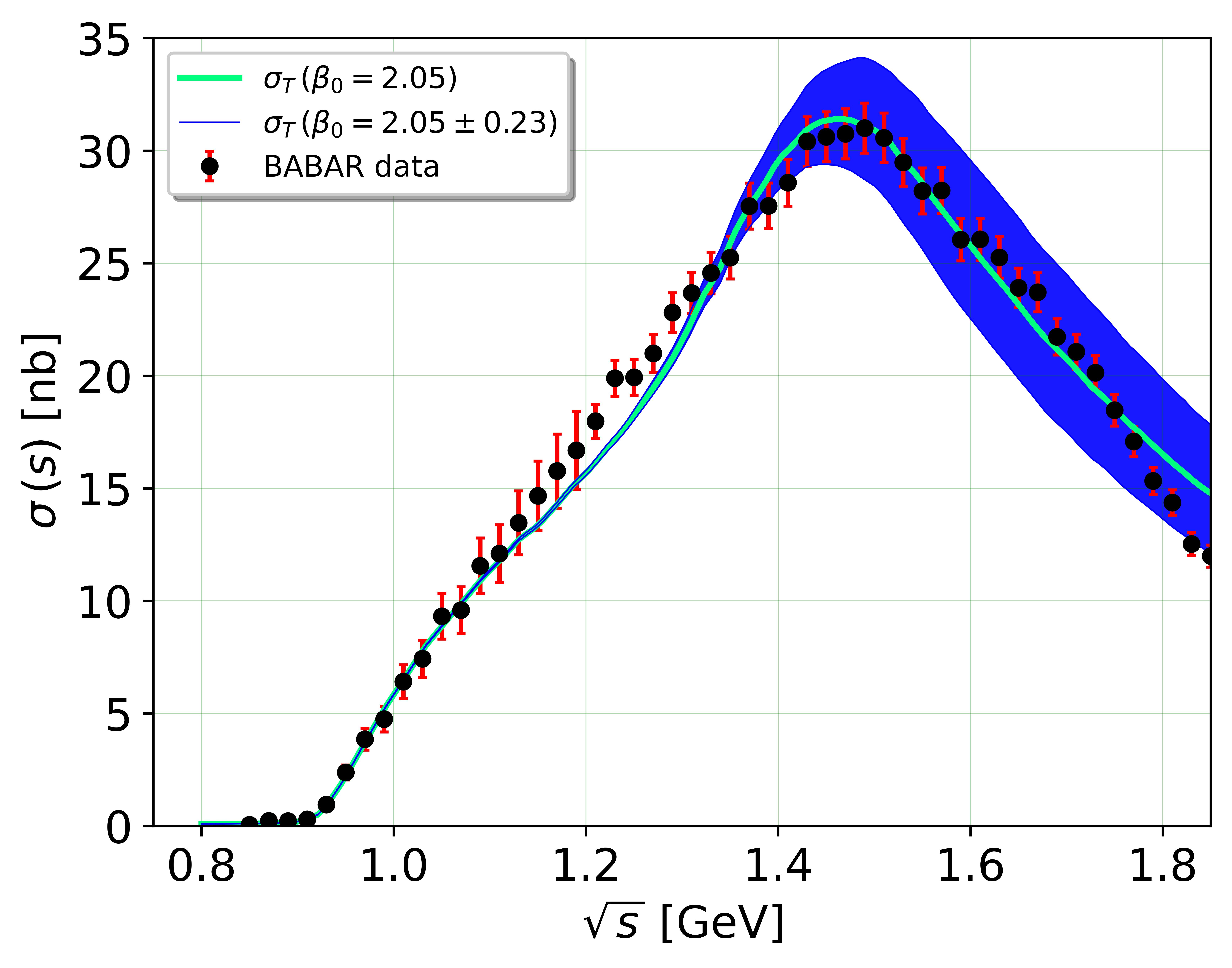}
\end{center}
\caption{Fit to the BaBar data for the total cross section $e^{+} e^{-} \to \pi^+ \pi^- 2 \pi^0$ with $\beta_0$ as the only free parameter. The blue band represents the corresponding uncertainty, which incorporates the uncertainty from the electric charge form factor normalization.}
\label{betacross}
\end{figure}

In Figure \ref{betacross}, we show the total cross section data from BaBar\cite{BaBar:2017zmc} (symbols)
and the fit (solid line), corresponding to $\beta_0=2.05$. The blue band represents the uncertainty, which incorporates the total uncertainty from the electric charge form factor normalization. The behavior is consistent with the previous observation in Figure \ref{betasense}, where the dependence on $\beta_0$ becomes important above 1.4 GeV. 

\section{Conclusions}
In order to determine the $\rho$ meson MDM, we have performed an analysis of the $e^+ e^- \to \pi^+ \pi^- 2 \pi^0$ cross section, measured by the BaBar Collaboration \cite{BaBar:2017zmc}. The $\gamma^* \to 4 \pi$  vertex was modeled in the VMD approach, including the exchange of the $\pi$, $\omega$, $a_1$, $\sigma$, $f_0(980)$, $\rho$ and $\rho'$ mesons. We obtained explicit gauge invariant expressions for the corresponding amplitudes, in particular for the $A$, $B$ and $C$ channels. This was done by properly combining the different contributions.
The model was tested in the low energy regime and in the $\omega$ channel, for which experimental data is provided \cite{BaBar:2017zmc}.
The best fit to the total cross section data implies a value for the MDM of the $\rho$ meson of $\mu_\rho=  2.7 \pm 0.3 $ in ($e/2 m_\rho)$ units, where the correction from the electric form factor introduced a large uncertainty, mainly due to the $\rho^\prime$ parameters. The different outcome with respect to the previous analysis \cite{GarciaGudino:2013alv,David2015} can be traced back to the difference in the handling of the $\rho^\prime$ parameters. Here, we profited from an analysis of a large number of observables \cite{Avalos:2022txl} to fix the model parameters which, although consistent with the previous work, shifted the central result to a higher value.
The theoretical description of the data is good enough to account for the $\rho$ MDM, at the given precision. We have exhibited the dependence on this parameter, and the missing structures observed in the data are not relevant for its determination.

\appendix

\section{Channels $A$, $B$ and $C$ gauge invariant amplitude}
We derive the fully gauge invariant amplitude for the $A$, $B$ and $C$ channels.
The gauge invariance condition requires:
\begin{equation}
q_{\mu}\,\Big(\mathcal{M}^{\mu}_{A} + \mathcal{M}^{\mu}_{B}\Big) = -q_{\mu}\,\mathcal{M}^{\mu}_{C}. 
\end{equation}
The amplitude $\mathcal{M}^{\mu}_{A}(p_1,p_2,p_3,p_4)$, Eq.~(\ref{MAr}), upon contraction with $q_{\mu}$ becomes:
\begin{equation}
    q_{\mu}\,\mathcal{M}^{\mu}_{A}(p_1,p_2,p_3,p_4) = i\,e\, \frac{g_{\rho\pi\pi}^{3}}{g_{\rho}}\,m^{2}_{\rho}\,D_{\rho^{0}}[q] \,z_{12}\cdot r_{43}\,D_{\rho^{-}}[s_{43}],\label{qMA1}
\end{equation}
where $x_1 \equiv q-2\,p_1$, $z_{12} \equiv q-p_1+p_2$ and $q=s_{21}+s_{43}$.\\
In a similar way $\mathcal{M}^{\mu}_{B}(p_1,p_2,p_3,p_4)$, Eq.~(\ref{MBr}), upon contraction with $q_{\mu}$, becomes:
\begin{eqnarray}
 &q_{\mu}\,{\cal{M}}^{\mu}_{B}(p_1,p_2,p_3,p_4) =
 i\,e\,\frac{g_{\rho\pi\pi}^{3}}{g_{\rho}}\,m^{2}_{\rho}\,D_{\rho^{0}}[q]\nonumber\\
 &\Big(r_{12}\cdot r_{43}\,D_{\rho^{-}}[s_{43}]-r_{43}\cdot r_{12}\,D_{\rho^{+}}[s_{21}]\Big).
\label{qMB1}
\end{eqnarray}

For channel $C$, Eq.~(\ref{MAr}), upon contraction with $q_{\mu}$ becomes:
\begin{eqnarray}
 q_\mu \,\mathcal{M}^{\mu}_{C}(p_1,p_2,p_3,p_4)&=&-i\,e\, \frac{g_{\rho\pi\pi}\,g_{\rho\rho\pi\pi}}{g_{\rho}}m^{2}_{\rho}\, D_{\rho^{0}}[q]\nonumber\\
&& q^\alpha \Gamma^{1}_{\alpha\gamma}\,D_{\rho^{-}}[s_{43}]\,r^{\gamma}_{43}.
\label{MrC1}
\end{eqnarray}

Choosing the following amplitudes combination, for $A$ and $B$ channels, we obtain:
\begin{eqnarray}
q_{\mu}\,\Big(\mathcal{M}^{\mu}_{A}(p_1,p_2,p_3,p_4) + \mathcal{M}^{\mu}_{A}(p_3,p_4,p_1,p_2) \nonumber\\
+\mathcal{M}^{\mu}_{B}(p_1,p_2,p_3,p_4)\Big)=
i\,e\,\frac{g_{\rho\pi\pi}g_{\rho\rho\pi\pi}}{g_{\rho}}m^{2}_{\rho}\, D_{\rho^{0}}[q]\nonumber\\
\Big\{\big(z_{12} - r_{21}\Big)\cdot r_{43}\,D_{\rho^{-}}[s_{43}]\nonumber\\
+\,\big(z_{34} -r_{43} \Big)\cdot r_{12}\,D_{\rho^{+}}[s_{21}]\Big\}.
\label{qMA14B1}
\end{eqnarray}

Choosing the following amplitudes combination, for C channel, we obtain: 
\begin{equation}
\begin{aligned}
 &-q_{\mu}\,\Big(\mathcal{M}^{\mu}_{C}(p_1,p_2,p_3,p_4) + \mathcal{M}^{\mu}_{C}(p_3,p_4,p_1,p_2)\Big)\\
 &\hspace{2 cm}=i\,e\,\frac{g_{\rho\pi\pi}\,g_{\rho\rho\pi\pi}}{g_{\rho}}\,m^{2}_{\rho}D_{\rho^{0}}[q] \\
 &\Big\{q^{\alpha}\,\Gamma^{1}_{\alpha\gamma}\,r^{\gamma}_{43}\,D_{\rho^{-}}[s_{43}] + q^{\alpha}\,\Gamma^{4}_{\alpha\gamma}\,r^{\gamma}_{12}\,D_{\rho^{+}}[s_{21}]\Big\}.\label{qMC14}
\end{aligned}
\end{equation}

Comparing Eq.~(\ref{qMA14B1}) and Eq.~(\ref{qMC14}), we identify $g_{\rho\rho\pi\pi} = g_{\rho\pi\pi}^{2}$ (for the case involving the $\rho^\prime$, the identification is $g_{\rho^{\prime}\rho\pi\pi} = g_{\rho^{\prime}\pi\pi}\,g_{\rho\pi\pi}$) and the structures are
\begin{equation}
 \begin{split}
  q^{\alpha}\,\Gamma^{1}_{\alpha\gamma} &= z_{12\gamma} - r_{21\gamma},\\
  q^{\alpha}\,\Gamma^{4}_{\alpha\gamma} &= z_{34\gamma} - r_{43\gamma}.
 \end{split}
\end{equation}
Factorizing $q^{\alpha}$, the $\Gamma^{i}_{\alpha\gamma}$ structures are built as:
\begin{equation}
 \begin{aligned}
 \Gamma^{1}_{\alpha\gamma} &=\Big\{z_{12\gamma}\,\frac{x_{3\alpha}}{x_3\cdot q}- r_{21\gamma}\,\frac{x_{3\alpha}}{x_3\cdot q} \Big\},\\
 \Gamma^{4}_{\alpha\gamma} &=\Big\{z_{34\gamma}\,\frac{x_{1\alpha}}{x_1\cdot q} - r_{43\gamma}\,\frac{x_{1\alpha}}{x_1\cdot q} \Big\}.
 \end{aligned}
\end{equation}
Thus, the final expression for the selected amplitudes of channel $C$ is:
\begin{align}
 &\mathcal{M}^{\mu}_{C}(p_1,p_2,p_3,p_4) = -i\,e\,\frac{g_{\rho\pi\pi}^3}{g_{\rho}}m^{2}_{\rho^{0}}\, D_{\rho^{0}}[q]\nonumber\\ 
 &\Big\{D_{\rho^{-}}[s_{43}]\,r_{43}\cdot z_{12}\,\frac{x_3{}^{\mu}}{x_3\cdot q} 
 +D_{\rho^{-}}[s_{43}]\,r_{43}\cdot r_{12}\,\frac{x_3{}^{\mu}}{x_3\cdot q} \Big\}.
\end{align}

 Thus, in order to obtain the gauge invariant amplitude, we have selected a particular set of amplitudes to combine among them, namely:

 \begin{eqnarray}
\mathcal{M}^{\mu}_{ABC_{24}} &=& \mathcal{M}^{\mu}_{A}(p_1,p_2,p_3,p_4)
+\mathcal{M}^{\mu}_{A}(p_3,p_4,p_1,p_2)\nonumber\\
&&+\mathcal{M}^{\mu}_{B}(p_1,p_2,p_3,p_4)\\
&&+\mathcal{M}^{\mu}_{C}(p_1,p_2,p_3,p_4)+\mathcal{M}^{\mu}_{C}(p_3,p_4,p_1,p_2).\nonumber
\end{eqnarray}

The final gauge invariant amplitude for this set of diagrams is given by Eq. (\ref{abc24}).
A similar expression corresponds to the $p_2 \leftrightarrow p_4$ exchange, adding the proper amplitudes Eq.~(\ref{abc42}) and $Q_{1} \equiv s_{43}-s_{21} \to Q_{2} \equiv s_{23}-s_{41}$.

\begin{acknowledgments}
We thank Pablo Roig for very useful observations. We acknowledge the support of CONAHCyT, Mexico Grant No. 711019 (A. R.) and the support of DGAPA-PAPIIT UNAM, under Grant No. IN110622. 
\end{acknowledgments}

\bibliography{rhomdm}% Produces the bibliography via BibTeX.

\end{document}